# A framework to decipher the genetic architecture of combinations of complex diseases: Applications in cardiovascular medicine

Liangying Yin[1], Carlos Kwan-long Chau[1], Yu-Ping Lin[1], Shitao Rao[1], Pak-Chung Sham[8], Hon-Cheong So[1-7*]

[1] School of Biomedical Sciences, The Chinese University of Hong Kong, Shatin, Hong Kong

[2] KIZ-CUHK Joint Laboratory of Bioresources and Molecular Research of Common Diseases, Kunming Institute of Zoology and The Chinese University of Hong Kong, China

[3] Department of Psychiatry, The Chinese University of Hong Kong, Hong Kong

[4] CUHK Shenzhen Research Institute, Shenzhen, China

[5] Margaret K.L. Cheung Research Centre for Management of Parkinsonism, The Chinese University of Hong Kong, Shatin, Hong Kong

[6] Brain and Mind Institute, The Chinese University of Hong Kong, Hong Kong SAR, China

[7] Hong Kong Branch of the Chinese Academy of Sciences Center for Excellence in Animal Evolution and Genetics, The Chinese University of Hong Kong, Hong Kong SAR, China

[8] Department of Psychiatry, University of Hong Kong, Hong Kong

**Correspondence to: Hon-Cheong So**, Lo Kwee-Seong Integrated Biomedical Sciences Building, The Chinese University of Hong Kong, Shatin, Hong Kong. Tel: +852 3943 9255; E-mail: hcso@cuhk.edu.hk

**Abstract**

Genome-wide association studies(GWAS) are highly useful in revealing the genetic basis of complex diseases. Currently, most GWAS are studies of a single disease diagnosis against controls. However, an individual is often affected by more than one condition. For example, patients with coronary artery disease(CAD) are often comorbid with diabetes(DM). Similarly, it is often clinically meaningful to study patients with one disease but *without* a related comorbidity. For example, obese DM may have different pathophysiology from non-obese DM.

Here we developed a statistical framework to uncover susceptibility variants for comorbid disorders(or a disorder without comorbidity), using GWAS summary statistics only. In essence, we mimicked a case-control GWAS in which the cases are affected with comorbidities or a disease without comorbidity(we may consider the cases as those affected by a specific disease 'subtype', as characterized by the presence/absence of comorbid conditions). We extended our methodology to analyze continuous traits with clinically meaningful categories(e.g. lipids). We also illustrated how the framework may be extended to more than two traits. We verified the feasibility and validity of our method by applying it to simulated scenarios and four cardiometabolic(CM) traits. We also analyzed the genes, pathways, cell-types/tissues involved in CM disease 'subtypes'. Genetic correlation analysis revealed that some subtypes may be biologically distinct from others. Further Mendelian randomization analysis showed differential causal effects of different subtypes to relevant complications. We believe the findings are of both scientific and clinical value, and the proposed method may open a new avenue to analyzing GWAS data.

**Introduction**

Genome-wide association studies have proven to be a highly useful design in revealing the genetic basis of many complex diseases, and has contributed to the understanding of the mechanisms of many diseases, for example in cardiovascular medicine and psychiatry[1,2]. GWAS data also have the potential to be directly translated to clinical practice, for example in risk prediction by polygenic scores, disease subtyping and drug discovery[3-8].

To date, more than 4000 GWAS have been conducted to date (https://www.ebi.ac.uk/gwas/), and the emergence of large biobanks (such as the UK Biobank) has further boosted the variety and amount of genomic data available. Most of the GWAS (or human sequencing studies) are studies of a particular *single* disease against controls. However, in clinical practice, an individual patient often has more than one condition/disorder. For example, patients with coronary artery disease (CAD) are often comorbid with diabetes mellitus (DM), while DM patients often have obesity; in psychiatry, patients with schizophrenia have a higher probability of having comorbid substance abuse, depression, obsessive compulsive disorder and many other psychiatric disorders[9].

Patients with *both* DM and CAD, for example, might share different pathophysiology than patients with DM *alone* or CAD *alone*. Viewed in another way, patients with both DM and CAD may be considered a 'distinct' entity, and its etiology and genetic basis may warrant further investigations. Ideally, we would perform a GWAS with 'cases' defined as patients having both disorders, and compared their genotypes with control subjects. However, recruiting patients with both disorders is usually more costly than recruiting those with a single disorder.

Along a similar line, it is often clinically meaningful to study patients with one disease but *without* a comorbid condition. For example, more than 90% of DM patients are overweight[10]; however there are still DM patients with normal weight, who may represent a specific *subtype* of DM. In the ideal case, we will wish to recruit patients with DM but normal weight as 'cases' and compare them against controls. However, the complexity and cost of recruitment is then increased (compared to the standard design of DM vs controls), hence limiting the practicality of such studies. As another example, it was estimated that ~75% of patients with depression also suffer from anxiety disorders[11]; however, the rest of the patients having depression but no anxiety disorders may represent a specific 'subtype' of depression. By studying the genetic basis of such subgroup of patients, we will be able gain deeper understanding into the heterogeneity and pathophysiology of depression.

As raised in the above examples, many diseases are highly heterogeneous. Patients with the same diagnosis may have different clinical presentations and prognosis, and share different etiologies. Through stratifying patients of the same diagnosis by the presence or absence of comorbid condition(s) and uncovering the genetic basis for each subgroup, we may gain better insight into the pathophysiology of the disorder. This will ultimately lead to more targeted interventions or prevention strategies for distinct subgroups of patients with the same diagnosis.

The aim of this study is to develop a framework to decipher the genetic architecture of multiple diseases *in combination*. Specifically, we wish to uncover susceptibility variants for comorbid conditions, and for 'subtype' of a disease without comorbid condition(s) (e.g. DM without obesity/overweight, depression without anxiety, CAD without hyperlipidemia etc.). The framework can be potentially applied to any complex diseases, and only summary statistics are required, which greatly extends the applicability of the methodology. In essence, we are 'mimicking' a case-control GWAS in which cases are affected with comorbidities, or affected by a particular disease but without a relevant comorbid condition (in either case, we may consider cases as having a 'subtype' of the disease as characterized by the presence or absence of comorbidities).

We will then apply such methods mainly to cardiovascular disorders (CVD), uncover genes and pathways associated, and uncover causal clinical risk factors for comorbidities (or disease without comorbidity). This study is mainly focused on applications in CVD in view of its high public health importance[12] and that many CVD are related to each other; nevertheless, the method itself is widely applicable to any complex traits. The presented computational framework can be considered an extension of the method by Nieuwboer et al.[13], for which the main focus was on finding susceptibility variants for functions of quantitative phenotypes such as body mass index (= weight/height$^2$). Here we modified and further developed the approach to accommodate binary outcomes, which are more commonly studied in GWAS, and proposed new applications in deciphering the genetic basis of disease subtypes as characterized by the presence (or absence) of a related trait. In addition, we also developed new methods to handle clinically defined categories of quantitative traits, approaches to compute covariance between phenotypes (which are required as input) as well as more general extensions to more than two diseases/traits.

Here we highlight a few related directions of research. One related research area is the finding of genetic correlation between complex traits. LD score regression (LDSR) is a commonly used technique to compute genetic correlation between traits[14] . Another related approach is to construct polygenic risk scores (PRS) for the first trait, and then test associations with the second. There is a fundamental difference between LDSR or PRS with the approach presented here. LDSR/PRS aims to discover *overall* genetic correlation or overlap between disorders, but are *not designed for finding specific susceptibility variants* underlying comorbid disorders or a disorder without comorbidity.

Another intuitive approach is to find variants passing a significance threshold (e.g. p<5e-8) for each trait, and directly find the overlapping variants. However, the setting of the significance threshold could be arbitrary, as setting a very stringent threshold (such as the conventional genome-wide significance cut-off) will lead to low power and carried the risk of missing genuine genetic variants contributing to comorbidity; setting a relaxed threshold (e.g. p<0.05) will result in increased false positive rates. Another approach to develop more formal statistical procedures to find shared genetic loci. For example, a co-localization approach based on summary statistics was proposed in Giambartolomei's paper[15]. The approach is Bayesian and outputs posterior probabilities that the variant is a genuine association signal for *both* traits. A limitation is that prior probabilities for different configurations of associations need to be specified, which may not be straightforward; difficulties may also arise for multiple independent associations at one locus. However, there are also *fundamental differences* between the 'co-localization' approach and our methodology. Our methodology can be conceptualized as mimicking the GWAS of a case-control study in which the cases are affected with comorbid disorders (or disease without a comorbid trait); as such, we are able to derive *effect sizes* [e.g. odds ratios(OR)] of individual genetic variants, and to conduct further analyses such as polygenic score analysis and Mendelian randomization. We may conclude, for example, the allele A (compared to a) of a certain SNP confers an OR of 2.0 to comorbid diabetes and heart disease. On the other hand, we cannot derive effect sizes with the co-localization approach. Also, the presented method is based on the frequentist approach with p-values as measures of significance, which may appear more familiar to many biologists and clinicians. Since most GWAS analytic tools are developed based on frequentist methods or use p-values as input, it might be easier to perform secondary analysis (such as gene- and pathway-based analysis) with our methodology.

GWAS meta-analysis is another related topic. However, the principle of meta-analysis is different from the proposed approach which address genetic basis of *combinations* of diseases/traits. In a meta-analysis, we aggregate evidence from different studies of the same or highly related traits to improve power; if one variant is highly significant in one large study, the final meta-analysis result will likely still be significant, regardless of the results of other studies. In addition, meta-analyses are not designed for finding genetic variants for a disease without a comorbid condition, such as DM with no obesity[16].

**Method**

In this study we introduce a statistical framework which has the potential to uncover susceptibility loci for comorbid disorders (or a disorder without comorbidity). It allows one to approximate the GWAS statistics for a comorbidity or a single disorder without comorbid trait based on the GWAS summary statistics of corresponding disorders only. R code for implementing the methodology is available at https://github.com/LiangyingYin/Infer_Summary_Statistics_for_Comorbidity_and_Single_disorder.

Here, we start by providing the derivation of GWAS summary statistics of interested trait (either a comorbid disorder or only single disorder without comorbidity) based on individual summary statistics. Suppose $P_1$ and $P_2$ are two different binary clinical traits. The interested clinical trait $T = f(P_1, P_2)$ can be defined as a function of the corresponding phenotypes. The following descriptions largely follows that of Nieuwboer et al.[13]; an extension to model comorbidities or disease without comorbid conditions is presented in the *next section*. Let $S \sim bin(n = 2, q)$ be a binomially distributed variable corresponding to the number of effect alleles (EA) of a biallelic SNP, where $q$ denotes the effect allele frequency. Suppose we have a multivariate linear regression model on a data set of size $N$, then we have:

$$\begin{bmatrix} P_{11} & P_{21} \\ P_{11} & P_{22} \\ \vdots & \vdots \\ P_{1N} & P_{2N} \end{bmatrix} = \begin{bmatrix} 1 & S_1 \\ 1 & S_2 \\ \vdots & \vdots \\ 1 & S_N \end{bmatrix} \begin{bmatrix} \beta_{01} & \beta_{02} \\ \beta_{11} & \beta_{12} \end{bmatrix} + \epsilon \qquad 2.1$$

where $P$ is a $N \times 2$ phenotype matrix, $S$ is a $N \times 2$ SNP matrix, $\beta$ is a $2 \times 2$ association estimates matrix, and $\epsilon$ is a $N \times 2$ error matrix. We assume each row $\epsilon$ is independent and follows a multivariate normal distribution $\epsilon \sim N(0, \Sigma)$. Our goal is to estimate $\gamma_0, \gamma_1$ for the target trait (either a comorbid disorder or only single disorder without comorbidity), i.e.,

$$f(P_{1i}, P_{2i}) =: T_i = \gamma_0 + \gamma_1 S_i + e_i \qquad 2.2$$

where $e_i$ follows a normal distribution with zero mean. It is equivalent to perform a GWAS for trait $T$. To realize this, we use the second-order Taylor approximation of $T$ around the point $\epsilon(s)$ for $s = 0, 1, 2$ where $\epsilon(s) := (E[P_{1i}|S_i = s], [P_{2i}|S_i = s])$. Here point $\epsilon(s)$ corresponds to the mean of the phenotypes of the individuals who has $s$ effect alleles for this SNP. The second-order Taylor approximation for the trait can be expressed as:

$$L_i := f(\epsilon(s)) + \sum_{k=1}^{2} \frac{\partial f(\epsilon(s))}{\partial P_k}(P_{ki} - E[P_{ki}|S_i = s]) + \frac{1}{2}\sum_{l=1}^{2}\sum_{k=1}^{2} \frac{\partial^2 f(\epsilon(s))}{\partial P_l \partial P_k}(P_{li} - E[P_{li}|S_i = s])(P_{ki} - E[P_{ki}|S_i = s]) \qquad 2.3$$

where $\partial f(\epsilon(s))/\partial P_k$ and $\partial^2 f(\epsilon(s))/\partial P_l \partial P_k$ denote the first- and second-order partial derivates of $f$ with respect to the corresponding phenotypes respectively, calculated at point $\epsilon(s)$. Conditioned on the allele count and taking expectation,

$$E[L_i|S_i = s] := E[f(\epsilon(s))|S_i = s] + E\left[\frac{1}{2}\sum_{l=1}^{2}\sum_{k=1}^{2}\frac{\partial^2 f(\epsilon(s))}{\partial P_l \partial P_k}(P_{li} - E[P_{li}|S_i = s])(P_{ki} - E[P_{ki}|S_i = s])|S_i = s\right] \qquad 2.4$$

as the second term in 2.3 becomes zero based on arguments in [13]. For simplicity and ease of exposition, if we ignore the quadratic or higher-order terms, a simple expression can be derived:

$$E[L_i|S_i = s] = \gamma_0 + \gamma_1 s \qquad 2.5$$

Then we will have a direct approximation for $\gamma_0$ when $s = 0$. Assuming we are studying two traits,

$$\hat{\gamma}_0 = E[L_i|S_i = 0] = f(\beta_{01}, \beta_{02}) \qquad 2.6$$

We can also estimate $\gamma_1$ by evaluate it for $s = 1, 2$ and weighting the results by their relative population frequency, i.e.,

$$\hat{\gamma}_1 = \frac{2q(1-q)}{2q(1-q)+q^2}(f(\beta_{01} + \beta_{11}, \beta_{02} + \beta_{12}) - \hat{\gamma}_0)$$
$$+ \frac{q^2}{2q(1-q)+q^2} \frac{(f(\beta_{01}+2\beta_{11},\beta_{02}+2\beta_{12})-\hat{\gamma}_0)}{2} \qquad 2.7$$

In practice one can include higher order terms in the approximation as needed. Since we do not have the covariance matrix of $\hat{\beta}$, we need to estimate it between each of the $\widehat{\beta_{ij}}$. Based on our multivariate linear regression assumption, the corresponding covariance matrix of $\hat{\beta}$ can be given by:

$$Var(\hat{\beta}) = (S^T S)^{-1} \otimes \Sigma \qquad 2.8$$

where $\Sigma$ is a $2 \times 2$ matrix with $\Sigma_{lk} = Cov(\epsilon_l, \epsilon_k)$, which is the covariance between errors in the linear regression of phenotypes $P_l, P_k$ on SNP S. We assume each individual SNP has small effect on the corresponding phenotype, so $Var(\epsilon_l) \approx Var(P_l)$ and $Cov(\epsilon_l, \epsilon_k) \approx Cov(P_l, P_k)$. Thus, we can infer

$$Cov(\widehat{\beta_{1l}}, \widehat{\beta_{1k}}) \approx SE_l Cor(P_l, P_k) SE_k \qquad 2.9$$

If there is only partial sample overlap, $Cov(\widehat{\beta_{1l}}, \widehat{\beta_{1k}})$ can be approximated as:

$$Cov(\widehat{\beta_{1l}}, \widehat{\beta_{1k}}) = SE_l Cor(P_l, P_k) \frac{N_{\cap l,k}}{\sqrt{N_l N_k}} SE_k \qquad 2.10$$

Where $N_l$ and $N_k$ are the number of individuals for the GWAS of $P_l$ and $P_k$ respectively, while $N_{\cap l,k}$ is the number of individuals present in both GWAS of $P_l$ and $P_k$. If $Cor(P_l, P_k)$ cannot be directly calculated, we can use LD score regression[14] to estimate $Cor(P_l, P_k) \frac{N_{\cap l,k}}{\sqrt{N_l N_k}}$ based on GWAS summary statistics. Notably, if there is no sample overlap between the phenotypes, then $N_{\cap l,k}$ is zero, so is the term $Cov(\widehat{\beta_{1l}}, \widehat{\beta_{1k}})$.

The covariance between intercepts $Cov(\widehat{\beta_{0l}}, \widehat{\beta_{0k}})$ can be expressed as:

$$Cov(\widehat{\beta_{0l}}, \widehat{\beta_{0k}}) = Cov(\overline{P_l} - \bar{S}\widehat{\beta_{1l}}, \overline{P_k} - \bar{S}\widehat{\beta_{1k}}) = Cov(\overline{P_l}, \overline{P_k}) - \bar{S}Cov(\widehat{\beta_{1l}}, \overline{P_k}) - \bar{S}Cov(\widehat{\beta_{1k}}, \overline{P_l}) + \bar{S}^2 Cov(\widehat{\beta_{1l}}, \widehat{\beta_{1k}}) \qquad 2.11$$

Since $Cov(\widehat{\beta_{1l}}, \overline{P_k})$ is zero and $Cov(\overline{P_l}, \overline{P_k})$ is negligible when sample size is large (typical GWAS sample sizes are >10,000), the above equation can be simplified as:

$$Cov(\widehat{\beta_{0l}}, \widehat{\beta_{0k}}) \approx \bar{S}^2 Cov(\widehat{\beta_{1l}}, \widehat{\beta_{1k}}) \qquad 2.12$$

Similarly, we may get

$$Cov(\widehat{\beta_{0l}}, \widehat{\beta_{1k}}) = Cov(\widehat{\beta_{1l}}, \widehat{\beta_{0k}}) = \bar{S}Cov(\widehat{\beta_{1l}}, \widehat{\beta_{1k}}) \qquad 2.13$$

**A framework for application to binary phenotypes - uncovering the genetic basis of comorbid disorders or a single disorder without related comorbidity**

The above derivations are based on continuous phenotypes. However, we are interested in disease traits which are usually binary. In this regard, we present a framework to deal with binary phenotypes and comorbid disorders or 'subtypes' of diseases without a comorbidity.

*Conversion of coefficient (from logistic model to that under a linear model and vice versa)*

In the above derivations it is assumed that we are dealing with coefficients obtained under a linear model. However, summary statistics for binary traits are usually derived from logistic regression. We therefore need to convert the coefficients from logistic models to those derived under linear models.

Lloyd-Jones et al[17] proposed a method for transforming summary statistics based on linear regression to odds ratio($OR_1$):

$$OR_1 = \frac{[k + \beta_1(1-p)][1 - k + \beta_1 p]}{[k - \beta_1 p][1 - k - \beta_1(1-p)]} \qquad 2.14$$

where $p$ indicates the effect allele frequency of the SNP under study $S$, $k$ represents the proportion of cases and $\beta_1$ represents the coefficient under a linear model. This formula is useful for converting coefficients from a linear model to those under a logistic model and vice versa (see below). Note that the odds ratio [=exp($\beta$)] estimate from a logistic regression is unbiased regardless of a retrospective or prospective design, with any level of over- or under-sampling of cases. This property however does not apply for linear regression[18]. To ensure that the final effect size estimate is close to the actual estimate when a prospective study is performed, we shall use the population lifetime risk estimate for $k$. Intuitively the analysis is performed as if we were doing a prospective study in the population. In the final step we will convert the coefficient from a linear model back to a logistic coefficient, which we shall employ the lifetime time probability of the comorbidity [i.e. Pr($P_1$ and $P_2$)] as input for $k$.

As explained above, here we are interested in the reverse of 2.14, i.e. solving for $\beta_1$ (coefficient under a linear model) given the odds ratio (OR). Denoting the odds ratio (OR) of SNP $S$ regressed on the binary phenotype by α, we have

$$(\alpha k - k)(1-k) = \beta_1^2[p(1-p) + \alpha p(1-p)] + \beta_1[\alpha k(1-p) + \alpha p - \alpha pk + kp + (1-p)(1-k)] \qquad 2.15$$

We could solve the above quadratic equation for $\beta_1$. We choose the solution whose absolute value is smaller than the coefficient under a logistic model, i.e., $abs(\beta_1) < abs(\log(\alpha))$. We verified this choice by experimenting with different $k$ and randomly generated $p$ from uniform distribution with value range of [0.05,0.95]. As demonstrated by Fig.S1, the absolute values of coefficient (abs($\beta$)) for binary traits derived from a linear regression is smaller than the absolute values of corresponding coefficients obtained under a logistic model [abs(log(α))] i.e., $abs(\beta) < abs(\log(\alpha))$.

After the conversion, we may compute $\widehat{\gamma_0}$ and $\widehat{\gamma_1}$. We can employ the delta method[19] to calculate the standard errors of $\widehat{\gamma_0}$ and $\widehat{\gamma_1}$. In essence, the delta method can be used to quantify variance of a function based on its first-order Taylor approximation. Suppose $G(X)$ and $U$ are the transform functions and mean vector of random variables $X = (x_1, x_2, \ldots x_n)$. The first-order Taylor expansion approximation for the function can be written as:

$$G(X) \approx G(U) + \nabla G(U)^T \cdot G(X - U) \qquad 2.16$$

where $\nabla G(U)$ is the gradient of $G(X)$. Then, we can take the variance of this approximation as the estimation for the variance of $G(X)$, i.e.,

$$Var(G(X)) \approx +\nabla G(U)^T \cdot Cov(X) \cdot \nabla G(X) \qquad 2.17$$

Covariance between the coefficients can be derived based on the methods described above.

*Modeling comorbid disorders or a disorder without comorbidity*

Let $P_1$ and $P_2$ be two different binary clinical traits (coded as 0 and 1 for the absence and presence of disease respectively), the presence of comorbidity (*Comor*) can be expressed as:

$$Comor = f(P_1, P_2) = P_1 \times P_2 \qquad 2.18$$

Thus we can infer the corresponding coefficient estimates as follows:

$$Comor(\widehat{\gamma_0}) = \widehat{\beta_{01}} \times \widehat{\beta_{02}} + Cov(P_1, P_2) \qquad 2.19$$

$$Comor(\widehat{\gamma_1}) = \frac{2q(1-q)}{2q(1-q)+q^2}\left[\left((\widehat{\beta_{01}} + \widehat{\beta_{11}}) \times (\widehat{\beta_{02}} + \widehat{\beta_{12}}) + Cov(P_1,P_2)\right) - Comor(\widehat{\gamma_0})\right] \qquad 2.20$$

$$+ \frac{q^2}{2q(1-q)+q^2} \frac{\left[\left((\widehat{\beta_{01}} + 2\widehat{\beta_{11}}) \times (\widehat{\beta_{02}} + 2\widehat{\beta_{12}}) + Cov(P_1,P_2)\right) - Comor(\widehat{\gamma_0})\right]}{2}$$

Here we also include the quadratic term (in 2.4) in the approximation. Similarly, having a disorder ($P_1$) but without a specific comorbidity ($P_2$) (e.g. CAD without DM) can be expressed as:

$$Single = f(P_1, P_2) = P_1 \times (1 - P_2) \qquad 2.21$$

And the corresponding coefficient estimates can be estimated by:

$$Single(\widehat{\gamma_0}) = \widehat{\beta_{01}} \times (1 - \widehat{\beta_{02}}) - Cov(P_1, P_2) \qquad 2.22$$

$$Single(\widehat{\gamma_1}) = \frac{2q(1-q)}{2q(1-q)+q^2}\left[\left((\widehat{\beta_{01}} + \widehat{\beta_{11}}) \times (1 - \widehat{\beta_{02}} - \widehat{\beta_{12}}) - Cov(P_1,P_2)\right) - Single(\widehat{\gamma_0})\right] \qquad 2.23$$

$$+ \frac{q^2}{2q(1-q)+q^2} \frac{\left[\left((\widehat{\beta_{01}} + 2\widehat{\beta_{11}}) \times (1 - \widehat{\beta_{02}} - 2\widehat{\beta_{12}}) - Cov(P_1,P_2)\right) - Single(\widehat{\gamma_0})\right]}{2}$$

Note that $cov(P_1, P_2)$ will cancel out in 2.20 and 2.23 hence this quantity does not affect estimates of $\widehat{\gamma_1}$.

*Extension to more than two traits*

The above proposed framework could also be extended to more than two traits. The only difference is that one of inputs should be the summary statistics of the comorbidity instead of a single disease, which could be derived from our proposed framework. For example, if we are interested in the genetic architecture of CAD comorbid with T2DM and obesity, we could estimate the summary statistics of *CAD comorbid with T2DM* first. Given the results, we could re-apply our methodology again to deal with comorbidity with the 3$^{rd}$ trait (obesity). It may be difficult to extract the lifetime risk of more than 2 comorbid disorders from the literature. If this is the case, we could employ Mendelian randomization (MR) to infer the OR of comorbid 1$^{st}$ + 2$^{nd}$ trait on the third one first, then the overall lifetime risk can be computed based on the methodology described in a section below. In a similar vein, the proposed framework can be applied to an arbitrary number of traits by *sequential* application of our method.

*Application to clinically defined categories of quantitative traits*

Our proposed framework is applicable to clinically defined categories of quantitative traits. For example, hyper- or dyslipidemia is a risk factor for many cardiovascular diseases and clinical thresholds for LDL-C, HDL-C and triglycerides have been defined to facilitate the identification and treatment for subjects at high risks. However, GWAS summary data are only available for lipids as a quantitative trait. One may wish to identify genetic variants contributing to for example comorbid CAD and hyperlipidemia, which is clinically relevant. To realize this, we need to transform coefficients derived from linear regression (with outcome as a continuous trait) ($\beta_1$) to coefficients from logistic regression (with outcome as a binary trait, such as hyperlipidemia or not) ($\beta_1^b$). This method assumes that the same genetic factors affect the quantitative trait across its distribution (same as the assumptions of a linear model on which the conversion is based); a limitation is that it may not discover genetic variants that have effects specifically in a certain range of the outcome distribution.

Suppose $S \sim bin(n = 2, q)$ is a binomially distributed SNP (where $q$ denotes the effect allele frequency,). For each continuous trait $y$, we may model the effects of each SNP by:
$$y = \beta_0 + \beta_1 S + \varepsilon \qquad 2.29$$
where ε is an error term that follows a normal distribution. The variance of the error term can be given by :
$$Var(\varepsilon) = Var(y) - \beta_1^2 Var(S) \qquad 2.30$$
Since individual SNP usually contributes to a very small explained variance, the residual variance of $y$ given $S$ is very close to the total variance of $y$. For each SNP $S$, we have:
$$E(S) = 2q(1-q) + 2q^2 \qquad 2.31$$
$$Var(S) = 2q(1-q) \qquad 2.32$$
Based on equation 2.29, we could compute the expected trait value for a given genotype, i.e.,
$$E(y|S = 0) = \beta_0 \qquad 2.33$$
$$E(y|S = 1) = \beta_0 + \beta_1 \qquad 2.34$$
$$E(y|S = 2) = \beta_0 + 2\beta_1 \qquad 2.35$$
Where $\beta_0$ can be calculated by :
$$\beta_0 = E(y) - \beta_1 E(S)$$
Given the genotype, the quantitative trait is assumed to follow a normal distribution. Since the mean [2.33 to 2.35] and variance [2.30, or approximated by *var(y)*] of the normal distribution are both known, we can infer the probability that the trait value will exceed (or fall below) certain threshold(s).

Then we could infer the corresponding odds for a clinically category (e.g. high LDL-C) for a given genotype, i.e., $odds_{S=0} = \frac{P(y|S=0)}{1-P(y|S=0)}$, $odds_{S=1} = \frac{P(y|S=1)}{1-P(y|S=1)}$ and $odds_{S=2} = \frac{P(y|S=2)}{1-P(y|S=2)}$. To estimate the coefficient $\beta_1^b$ when clinically defined categories of a quantitative trait are considered as the outcome, we could evaluate the above at $S = 1,2$ and weigh them by their relative population frequency, i.e.

$$\beta_1^b = \frac{2q(1-q)}{2q(1-q)+q^2} \log\left(\frac{odds_{S=1}}{odds_{S=0}}\right) + \frac{q^2}{2q(1-q)+q^2} \frac{\log\left(\frac{odds_{S=2}}{odds_{S=0}}\right)}{2} \qquad 2.39$$

We may then use the delta method to calculate standard error of $\beta_1^b$.

*Computing probability of comorbidity $P_1$ and $P_2$*

Our calculations require specifying the probability of comorbidity, $E(P_1 P_2)$, for conversion of the linear coefficient back to the logistic scale in the final step of the algorithm. This measure is also required when the methodology is to be applied to three or more traits. While estimates could be obtained from related literature, the lifetime probability for comorbidity could be relatively hard to find.

If this is the case, we can calculate it from the OR (or relative risk, RR) of trait $P_1$ given the other trait $P_2$ (in which $P_2$ can be considered a risk factor). For example, one may obtain the OR of CAD given diabetes based on literature search or other means. Based on Bayes rule, we have:

$$E(P_1 P_2) = \Pr(P_1 = 1 \text{ and } P_2 = 1) = \Pr(P_1 = 1|P_2 = 1) \times \Pr(P_2 = 1) \qquad 2.25$$

Here we shall develop an approach to compute $\Pr(P_1 = 1|P_2 = 1)$ given OR or RR. Let $f_{RF0}$ be the probability of having the disease ($P_1$) given the *absence* of the risk factor ($P_2$), $f_{RF1}$ be the probability of having the disease ($P_1$) given the *presence* of the risk factor ($P_2$), $P_{RF0}$ denote the probability of having no risk factor i.e. $Pr(P_2=0)$, and $P_{RF1}$ denote the probability of having the risk factor i.e. $Pr(P_2=1)$.

From 2.25, we may also express $E(P_1 P_2)$ as $f_{RF1}$ x $P_{RF1}$.

Note that we have $K = f_{RF0} P_{RF0} + f_{RF0} RR \cdot P_{RF1}$ or

$$f_{RF0} = K/(P_{RF0} + RR \cdot P_{RF1}) \qquad 2.26$$

When the RR of disease given the risk factor is available, the calculation is straightforward as by definition we have $f_{RF1} = RR * f_{RF0}$ and $E(P_1, P_2) = f_{RF1}$ x $P_{RF1}$. The case is more complicated when only OR are available. Here we present an iterative procedure to estimate $E(P_1 P_2)$. In the first step we may use OR to approximate RR,

$$f_{RF0} \approx K/(P_{RF0} + OR \cdot P_{RF1}) \qquad 2.27$$

From Zhang et al[20], OR may be estimated from RR by

$$RR \approx OR/(1 - f_{RF0} + f_{RF0} \cdot OR) \qquad 2.28$$

The newly estimated RR from 2.28 can be substituted back into 2.26 to obtain a new estimate of $f_{RF0}$. The algorithm is iterated until RR becomes stable (change in RR between iterations <1e-10). Finally we can compute the probability of comorbidity by $E(P_1, P_2) = f_{RF1} \times P_{RF1}$.

**Simulation study**

*Application to binary traits*

To verify the feasibility and validity of our proposed framework, we simulated different sets of genotype-phenotype data, with 300 SNPs (*i.e.* $N_{snp}$ = 300; coded as 0, 1, 2) and two binary traits. As the proposed framework is a SNP-based analysis, the number of simulated SNPs shall not affect the validity of our simulations. For each simulated SNP, the allele frequency was randomly generated from a uniform distribution with a value range of [0.05, 0.95]. The number of subjects with each disorder (i.e., $ncases$) was set to [10000, 20000, 50000, 100000] with a disease prevalence ($K$) of 10%. Here, $ncases$ indicated the sample size of cases in the whole simulated population cohort. Sample size of the whole population ($ntotal$) was estimated by $ntotal = \frac{ncases}{K}$. Then, based on the generated allele frequencies, we could infer the genetic profiles for the whole simulated population cohort. The total SNP-based heritability ($h^2$) for each trait was set at 0.02 to 0.3, distributed among all SNPs. More specifically, we simulated standard normal variables $z_i \sim N(0,1)$, and set mean effect size $\mu = \sqrt{\frac{h^2}{N_{snp}}}$. The actual effect size for SNP$_i$ was set at $\beta_i = \mu \times z_i$. The total liability $y$ equals the sum of effects from each SNP plus a residual ($e$), i.e. $y = \sum_i \beta_i x_i + e$; the total variance of $y$ was set to one. Following the liability threshold model, subjects with total liability exceeding a certain threshold [$= \Phi^{-1}(K)$, where $K$ is the disease prevalence] are regarded as being affected by the disease. The non-shared genetic covariance between the two traits was set to 0.01 and 0.1 respectively.

From the simulated population cohort, we simulated two case-control studies with traits A and B as the outcome respectively. Suppose the number of cases for trait A and B in the population are $N_A$ and $N_B$ respectively, and $N = max(N_A, N_B)$. For trait A, we picked $N_A$ cases and $2N - N_A$ controls from the population. For trait B, we picked $N_B$ cases and $2N - N_B$ controls from the population. For comparison, we also simulated a *'real'* comorbidity GWAS by picking all the comorbid cases ($N_{comor}$) in the simulated population cohort who are identified as being affected by *both* trait A and B. Then ($2N - N_{comor}$) controls were included.

Case-control samples for the two traits with different overlap rates (*P*) were simulated to demonstrate the feasibility of our proposed method. Here *P* was defined as the ratio of overlapped samples (including overlapped cases $N_{AB.overlap}$ and controls $N_{ctrl.overlap}$) to the total sample size (2*N*) for each case-control study, i.e., $P = (N_{AB.overlap} + N_{ctrl.overlap})/2N$. To adjust the overlap rate, we would increase or decrease the number of shared cases and/or controls for both traits. Two different overlap rates were simulated for both comorbid and single disorder. Most shared subjects were controls as they were far more abundant than cases (this is also likely the case in real applications).

We also considered another type of study design, namely a prospective study of a population. The simulation scheme is similar to the above, except that the controls consisted of the rest of the population who were not cases. Again, we assessed the performance of our proposed method under different overlap rates. We simulated completely and partially overlapping samples. Besides, we studied the performance of our proposed framework both for the case where all required parameters were given and for the case where disease prevalence was mis-specified.

*Transforming regression coefficients for quantitative traits to those for binary traits (based on clinically meaningful categories)*

We simulated datasets to verify the feasibility of our proposed approach to 'transform' regression coefficients of quantitative traits to coefficients based on clinically defined categories. We simulated a quantitative trait and derive analytic estimates of GWAS summary statistics of a *dichotomous* trait derived from the quantitative measure (e.g. hyperlipidemia). The 'real' GWAS scenario refers to directly considering the dichotomous trait as the outcome in regression.

The number of subjects for quantitative traits was set to [50000, 10000]. For each simulated SNP, the allele frequency was randomly generated from a uniform distribution from [0.05, 0.95]. SNP-heritability was set to 0.1 and 0.2. The aim was to compare the theoretical estimates of regression coefficients and SE (based on summary statistics alone) against those obtained from simulated raw genotype data.

**Application to binary traits in cardiovascular medicine**

We applied our approach to 4 cardiometabolic disorders/traits, namely coronary artery disease[21] (CAD), type 2 diabetes mellitus[22] (T2DM), obesity[23] (BMI>=30) and stroke[24] (all types of stroke included), based on publicly available GWAS summary statistics. Details of these datasets are summarized in Table S1. Since the effect size of individual SNPs in different GWAS may not correspond to the same allele, we employed the 'harmonise_data' function in the package TwoSampleMR (https://mrcieu.github.io/TwoSampleMR) which integrates GWAS summary statistics from different sources, taking into account DNA strand issues[25-27]. Analysis was performed for SNPs with MAF>=0.01. In total, we studied the genetic architectures of 18 disease 'subtypes' (6 are comorbidities and the remaining 12 are 'single' diseases without relevant comorbid conditions) and identified contributing genetic variants.

**Genes, pathway and cell type enrichment analysis**

To better understand the functional and biological mechanisms underlying the genetic component of these disease 'subtypes', we computed gene-based significance using MAGMA inserted in the web-based tool FUMA[28,29]. We employed false discovery rate (FDR) for multiple testing correction and selected genes that with FDR<0.05 for further analysis. Also, we performed "tissue specificity" analysis by examining whether the susceptibility genes are differentially expressed in a particular tissue. Apart from these analyses, we also conducted pathway analysis using the program "ConsensusPathDB", in order to unravel biological pathways that are unique to specific disease subtype and shared pathways among disease subtypes. Furthermore, we examined the cell types that are enriched for specific disease 'subtypes'.

**Finding heritability explained by common variants**

To further understand the genetic architecture of the 'subtypes' of complex diseases, we calculated their SNP-based heritability by LD score regression (LDSR). We also explored the genetic correlation between different disease subtypes with LDSR. Our aim is clarify whether disease "subtypes" are genetically different from each other.

**Mendelian Randomization (MR) analysis**

Mendelian Randomization (MR) is a methodology for inferring the causal relationship between risk factors and outcomes, using genetic variants as 'instruments' to represent the exposure. Here we performed two-sample MR in which the instrument-exposure and instrument-outcome associations were estimated in different samples. MR was conducted with 'inverse-variance weighted' (MR-IVW)[30] and Egger regression (MR-Egger)[25] approaches. MR-Egger is able to give valid estimates of causal effects in the presence of imbalanced horizontal pleiotropy; the latter was assessed by whether the Egger intercept was significantly different from zero.

The IVW framework is widely used in MR. Here we used an IVW approach that is able to account for SNP correlations[30]. A similar approach may be used for MR-Egger, which allows an intercept term in the weighted regression. Please refer to[25] and its Supplementary Text for details. Inclusion of a larger panel of SNPs in partial LD may enable higher variance to be explained, thus improving the power of MR[27]. Including "redundant" SNPs in addition to the causal variant(s) would not invalidate the results, although including too many variants with high correlations may result in unstable causal estimates[27].

To ensure the robustness of our findings, we performed MR at multiple $r^2$ thresholds (0.001, 0.01, 0.05, 0.1 and 0.15) with SNP correlations taken into account. For simplicity, we mainly present the results at $r^2$=0.05. However, our findings are consistent across various thresholds and full results are given in Table S10.

Only SNPs which passed genome-wide significance ($p$<5E-8) were included as instruments. We employed the R packages "MendelianRandomization" (ver 0.4.1) and "TwoSampleMR" (ver 4.25) for analysis. If a SNP was not available in the outcome GWAS, we allow using a "proxy SNP" provided $r^2$>=0.8 with the original SNP. LD was extracted from the 1000 Genomes European samples.

Multiple testing was corrected by the false discovery rate (FDR) approach by Benjamini and Hochberg.

**Replication in real data**

To further verify our proposed approach, we also calculated genetic correlations between GWAS results obtained from actual GWAS of comorbid/single disorders and those derived from our approach. Specifically, we firstly performed GWAS on CAD with T2DM, CAD without T2DM, and T2DM without CAD cases in the UK Biobank (UKBB). GWAS was conducted by Regenie[31], which was reported to be suitable for analyzing a large number of samples and multiple phenotypes simultaneously while accounting for relatedness and population structure. Briefly, the program was run in 2 steps, in which the 1$^{st}$ step employed a subset of genetic markers to fit a whole-genome regression model to capture phenotypic variance and the 2$^{nd}$ step tested associations of a larger set of variants with the outcome conditional on predictions from the model derived from the 1$^{st}$ step. The covariates included sex, age and top 10 principal components[32]. Quality control was performed prior to GWAS; briefly, subjects with missing data rates ≥10% were dropped from the analyses, and SNPs were dropped if they had either a minor allele frequency<1%, a missing data rate ≥10% or a Hardy-Weinberg equilibrium (HWE) p-value <10e-06. For UKBB, CAD and T2DM were defined by a combination of self-reported and ICD10-coded disease. We applied the proposed methodology to GWAS summary statistics from ref.[21,22], as described above.

**Results**

**Simulation results**

*Simulation results for binary traits*

Results of our simulation are presented in Tables 1 and 2. For more detailed simulation results, please refer to Tables S2 and S3. The correlations between the estimated and actual coefficients were in general very high. The correlation and RMSE improved with increased sample sizes and higher heritability explained by SNPs (i.e. with larger effect sizes of SNPs). Since current GWAS summary data are usually of very large sample sizes, often larger than 100,000, we believe the current method is sufficiently good to approximate the results for a GWAS of comorbidity or other combination of diseases/traits. Also, the current method is valid under different rates of overlap between the input GWAS datasets.

It is obvious that the power increased with larger samples sizes and heritability explained. The type I error is kept at or below 0.05 when the significance threshold is set at p<0.05. Interestingly, the power of the proposed analytic method is often higher than the simulated 'actual' GWAS with genotype data when we study comorbid disorders. This may be because only a small number of patients were affected with both diseases ($N_{comor}$ is low); on the other hand, the number of subjects affected with either disease is larger, therefore the two sets of case-control GWAS data (of traits A and B) may contain more information than a GWAS on the minority of individuals affected by both diseases, under the same total samples size. For a single disease without comorbidity, the power of the proposed method is close to that of the simulated 'actual' GWAS.

Table S4 demonstrates the simulation results with misspecified lifetime risk of disease for both comorbid and 'single' disorder. The proposed method is reasonably robust to misspecified lifetime risk of disease.

Table 1 Simulation results for comorbid disorders compared with "real" GWAS

| Overlap rate | | No. cases | $H^2$ A | $H^2$ B | Correlation | | RMSE | | Inferred | | Real GWAS | |
|---|---|---|---|---|---|---|---|---|---|---|---|---|
| Cases | Controls | | | | Beta | SE | Beta | SE | Power | Type I error | Power | Type I error |
| 0.08 | 0.15 | 10000 | 0.2 | 0.3 | 0.93024 | 0.89173 | 0.05778 | 0.02426 | 0.633 | 0.027 | 0.503 | 0.047 |
| | | 20000 | 0.2 | 0.3 | 0.95848 | 0.89268 | 0.04868 | 0.01752 | 0.753 | 0.05 | 0.620 | 0.033 |
| | | 50000 | 0.2 | 0.3 | 0.98365 | 0.88681 | 0.04131 | 0.01103 | 0.853 | 0.047 | 0.753 | 0.037 |
| | | 100000 | 0.2 | 0.3 | 0.99038 | 0.89388 | 0.03730 | 0.00776 | 0.87 | 0.04 | 0.793 | 0.05 |
| | | 10000 | 0.08 | 0.09 | 0.84612 | 0.95722 | 0.05011 | 0.02364 | 0.473 | ------ | 0.297 | ------ |
| | | 20000 | 0.08 | 0.09 | 0.91102 | 0.95360 | 0.03524 | 0.01696 | 0.603 | ------ | 0.443 | ------ |
| | | 50000 | 0.08 | 0.09 | 0.95823 | 0.95246 | 0.02483 | 0.01077 | 0.763 | ------ | 0.610 | ------ |
| | | 100000 | 0.08 | 0.09 | 0.98354 | 0.95457 | 0.01860 | 0.00760 | 0.807 | ------ | 0.703 | ------ |
| 0.04 | 0.4 | 10000 | 0.2 | 0.3 | 0.91931 | 0.88427 | 0.05787 | 0.02271 | 0.6 | 0.043 | 0.500 | 0.046 |
| | | 20000 | 0.2 | 0.3 | 0.96065 | 0.88222 | 0.04504 | 0.01619 | 0.743 | 0.04 | 0.617 | 0.043 |
| | | 50000 | 0.2 | 0.3 | 0.97973 | 0.88937 | 0.03877 | 0.01026 | 0.8 | 0.037 | 0.740 | 0.05 |
| | | 100000 | 0.2 | 0.3 | 0.98954 | 0.89279 | 0.03657 | 0.00726 | 0.89 | 0.043 | 0.837 | 0.04 |
| | | 10000 | 0.08 | 0.09 | 0.81907 | 0.94187 | 0.05180 | 0.02203 | 0.473 | ------ | 0.307 | ------ |
| | | 20000 | 0.08 | 0.09 | 0.90766 | 0.94663 | 0.03534 | 0.01577 | 0.583 | ------ | 0.420 | ------ |
| | | 50000 | 0.08 | 0.09 | 0.95481 | 0.94882 | 0.02663 | 0.00996 | 0.740 | ------ | 0.590 | ------ |
| | | 100000 | 0.08 | 0.09 | 0.97193 | 0.95743 | 0.02258 | 0.00703 | 0.800 | ------ | 0.707 | ------ |

Note: here No. cases indicates the number of cases we defined for our simulation scenarios, $H^2$ indicates heritability, RMSE is abbreviated for root mean square error. Please refer to the main text for details on simulation methods. The 'real' GWAS was constructed by $N_{comor}$ cases and $(2N - N_{comor})$ controls.

Table 2 Simulation results for only single disorder compared with "real" GWAS

| Overlap rate | | No. cases | $H^2$ A | $H^2$ B | Correlation | | RMSE | | Inferred | | Real GWAS | |
|---|---|---|---|---|---|---|---|---|---|---|---|---|
| Cases | Controls | | | | Beta | SE | Beta | SE | Power | Type I error | Power | Type I error |
| 0.13 | 0.15 | 10000 | 0.2 | 0.3 | 0.95485 | 0.99823 | 0.02883 | 0.00039 | 0.623 | 0.047 | 0.617 | 0.050 |
| | | 20000 | 0.2 | 0.3 | 0.97665 | 0.99827 | 0.02084 | 0.00036 | 0.723 | 0.050 | 0.713 | 0.040 |
| | | 50000 | 0.2 | 0.3 | 0.98305 | 0.99842 | 0.01742 | 0.00021 | 0.820 | 0.050 | 0.820 | 0.047 |
| | | 100000 | 0.2 | 0.3 | 0.98887 | 0.99848 | 0.01418 | 0.00014 | 0.867 | 0.050 | 0.877 | 0.020 |

| | | | | | | | | | | |
|---|---|---|---|---|---|---|---|---|---|---|
| | 10000 | 0.08 | 0.09 | 0.91332 | 0.99939 | 0.02590 | 0.00023 | 0.460 | ------ | 0.473 | ------ |
| | 20000 | 0.08 | 0.09 | 0.94697 | 0.99955 | 0.01946 | 0.00017 | 0.583 | ------ | 0.583 | ------ |
| | 50000 | 0.08 | 0.09 | 0.97378 | 0.99962 | 0.01333 | 0.00008 | 0.747 | ------ | 0.740 | ------ |
| | 100000 | 0.08 | 0.09 | 0.97938 | 0.99958 | 0.01174 | 0.00008 | 0.800 | ------ | 0.783 | ------ |
| | 10000 | 0.2 | 0.3 | 0.95604 | 0.99722 | 0.02918 | 0.00055 | 0.630 | 0.047 | 0.630 | 0.033 |
| | 20000 | 0.2 | 0.3 | 0.97607 | 0.99766 | 0.02101 | 0.00031 | 0.737 | 0.047 | 0.737 | 0.037 |
| | 50000 | 0.2 | 0.3 | 0.98656 | 0.99765 | 0.01567 | 0.0002 | 0.830 | 0.050 | 0.813 | 0.047 |
| 0.21 0.25 | 100000 | 0.2 | 0.3 | 0.99113 | 0.99781 | 0.01255 | 0.00014 | 0.880 | 0.043 | 0.883 | 0.047 |
| | 10000 | 0.08 | 0.09 | 0.92425 | 0.99897 | 0.02342 | 0.00054 | 0.430 | ------ | 0.433 | ------ |
| | 20000 | 0.08 | 0.09 | 0.95138 | 0.99917 | 0.01892 | 0.00030 | 0.607 | ------ | 0.570 | ------ |
| | 50000 | 0.08 | 0.09 | 0.97327 | 0.99931 | 0.01385 | 0.00020 | 0.737 | ------ | 0.743 | ------ |
| | 100000 | 0.08 | 0.09 | 0.979642 | 0.999344 | 0.011849 | 0.000130 | 0.783 | ------ | 0.780 | ------ |

Note: here No. cases indicates the number of cases we defined for our simulation scenarios, $H^2$ indicates heritability, RMSE is abbreviated for root mean square error.

*Simulation results for clinically defined categories of quantitative trait*

Table 3 shows the simulation results for clinically defined categories of quantitative traits. The correlations between the estimated and actual coefficients are high, and the correlation and RMSE improved with increased sample sizes and higher heritability explained by SNPs. The slightly higher power of using the analytic approach may be due to loss of information when dichotomizing the continuous trait in the simulated 'real' GWAS.

Table 3 Simulation results for clinically defined categories of a continuous trait

| Sample size | Case sample size | $H^2$ | Correlation | | RMSE | | Power | | Type I error | |
|---|---|---|---|---|---|---|---|---|---|---|
| | | | Beta | SE | Beta | SE | Inferred | Real GWAS | Inferred | Real GWAS |
| 50000 | 7635 | 0.1 | 0.96518 | 0.99905 | 0.01596 | 0.00701 | 0.657 | 0.497 | 0.023 | 0.033 |
| 50000 | 10336 | 0.1 | 0.97356 | 0.99936 | 0.01385 | 0.00612 | 0.657 | 0.507 | 0.023 | 0.027 |
| 50000 | 7719 | 0.2 | 0.98126 | 0.99825 | 0.01626 | 0.00707 | 0.783 | 0.627 | 0.023 | 0.033 |
| 50000 | 10455 | 0.2 | 0.98782 | 0.99894 | 0.01255 | 0.00545 | 0.783 | 0.653 | 0.023 | 0.023 |
| 100000 | 15789 | 0.1 | 0.98216 | 0.99909 | 0.01138 | 0.00503 | 0.727 | 0.607 | 0.043 | 0.030 |
| 100000 | 21103 | 0.1 | 0.98572 | 0.99966 | 0.00993 | 0.00417 | 0.727 | 0.650 | 0.047 | 0.063 |
| 100000 | 15773 | 0.2 | 0.99133 | 0.99825 | 0.01112 | 0.00503 | 0.817 | 0.727 | 0.043 | 0.030 |
| 100000 | 21345 | 0.2 | 0.99227 | 0.99894 | 0.01003 | 0.00387 | 0.817 | 0.773 | 0.043 | 0.063 |

Note: here Sample size indicates the sample size of our simulated dataset, Case size indicates number of cases based on our clinically defined categories, $H^2$ indicates heritability, RMSE is abbreviated for root mean square error.

**Application to cardiovascular disorders/traits**

The proposed framework was applied to 4 cardiometabolic (CM) diseases/traits, the combination of which results in 18 disease 'subtypes' (6 are comorbidities and the remaining 12 are 'single' diseases without relevant comorbid conditions). We estimated the effect size (in terms of odds ratio comparing subjects with the disease 'subtype' versus those without) and the corresponding SE and p-values based on our presented analytic framework. Following the definition by the GWAS analytic platform FUMA (https://fuma.ctglab.nl/tutorial#riskloci), independent significant SNPs are defined as those with $p<5e-8$ and independent from each other at the default $r^2$ threshold ($r^2=0.6$). As for the definition of genomic *loci*, independent significant SNPs which are correlated with each other at $r^2 \geq 0.1$ are assigned to the same risk locus. Independent significant SNPs which lie within 250 kb are also merged into one genomic risk locus. For the lifetime risks of the diseases under study, some of them were directly extracted from relevant literatures[33-38], while the remaining were inferred from ORs (or RRs) from relevant studies (see Table S5).

*Genes, cell type and pathway analysis*

Here we report the analysis results for the 18 disease 'subtypes'. In total, we identified 384 and 587 genomic risk loci respectively for 6 comorbidities and 12 disease 'subtypes' without a relevant comorbid condition (Table 4, Fig.1 and Table S6). Here we take Type 2 DM and obesity and the combination of these two traits as example. As expected, some susceptibility genes were shared among disease 'subtypes'. For instance, *TCF7L2* and *CDKAL1*[39] are the top susceptibility genes for all 3 disease subtypes involving T2DM and obesity[40]. Notably, *TCF7L2*[41-43] and *CDKAL1*[44,45] are also among the top genes for all three disease subtypes involving CAD and T2DM, suggesting a more general role of these genes in the pathogenesis of various forms of cardiometabolic abnormalities. Some susceptibility genes were only identified in specific disease subtypes. For example, *FTO* was found to be implicated only in disease subtypes that involved obesity, i.e., obesity with or without T2DM, but *not* T2DM without obesity. This finding is consistent with previous studies that *FTO* mainly contributes to diabetes through its effects on BMI[46-48]. Interestingly, *BDNF* was among the top genes for obese DM. BDNF treatment has been shown to reduce weight gain and glucose level in animal models and was also associated with glucose metabolism in clinical studies[49]. On the other hand, genes such as *JAZF1, HMGA2, COBLL1, KCNJ11* and *PPARG* were ranked among the top for *non-obese* DM, indicating these genes may contribute to glucose dysregulation other than through effects on BMI/obesity. Notably, *KCNJ11* and *PPARG* are also drug targets for sulphonylureas and thiazolidinediones (known anti-DM medications); further studies on the mechanisms and clinical efficacy of these classes of drugs in non-obese DM subjects may be warranted. For details about the concordant and discordant genes among disease subtypes, please refer to Table S7.

In addition, we also performed pathway analysis through the tool ConsensusPathDB. The enriched pathways were summarized in Table S8. Similar to gene analysis, some enriched pathways were shared among different disease subtypes while others were unique to particular disease subtype. Taking the 3 disease subtypes involving CAD and T2DM as example, statin pharmacodynamics, transcriptional regulation by RUNX3, and Angiopoietin receptor Tie2-mediated signaling were significantly enriched in all three disease subtypes, suggesting a boarder role of these biological pathways across CVD. There were also pathways that were only significantly enriched in certain disease subtype. For example, amb2 Integrin signaling[50] and chylomicron/plasma lipoprotein clearance[51] were top-ranked for *non*-diabetic CAD. As another example, adipogenesis and MAPK cascade pathways were enriched in CAD comorbid with T2DM. Previous studies have implicated a role of MAPK cascade in the pathogenesis of both cardiac diseases and diabetes[52-54]. By investigating the pathways enriched for each disease 'subtype', we hope to gain insight into biological mechanisms that are generally important across CM disorders, as well as more 'specific' mechanisms that may play a more salient role for certain disease combinations.

Next we also performed cell-type and tissue specificity analysis through FUMA. FUMA pre-computes a list of genes differentially expressed in different tissues (DEGs) from GTEx; input genes (significant genes from MAGMA analysis in GWAS) are then tested for enrichment for these DEGs. This approach is simple but we note that differential expression does not always suggest causal role of the tissue. We consider this as a hypothesis-generating analysis. According to the results, coronary artery and aorta were the most significantly enriched tissues only for disease subtypes that involved CAD. Interestingly, for disease subtypes including obesity without CAD, obesity without stroke and non-obese T2DM, the most significantly enriched tissues include brain tissues such as frontal cortex and cerebellum (Fig. S2). While the results will require further experimental validation, the brain has been suggested to play a key role in the control of body fat content and glucose metabolism[55,56].

Recently methods have been developed for cell-type enrichment analysis based on GWAS[57], as single-cell sequencing data becomes more widely available. However, single-cell data to date are more abundant for the brain than for other tissue types. This part is considered more exploratory as not all cell types are available for analysis in FUMA. We shall focus on the enrichment results for several common comorbidities (most other disease combinations did not return significant results). To highlight a few interesting findings (Fig. S3), we found GABAergic neurons in the midbrain and prefrontal cortex to be the most enriched cell type for CAD with obesity. Interestingly, it has been reported that leptin exerts its anti-obesity effects mostly through GABAergic neurons in the brain[58]. GABA neurotransmission is also thought to play a role in appetite regulation[59]. On the other hand, GABA in the CNS may also regulate the sympathetic outflow to the coronary vasculature, causing a change in vascular resistance[60]. For CAD with T2DM, we also found enrichment of endothelial cells in the pancreas.

Table 4 Genome-wide significant SNPs and risk loci for studied disease subtypes

| Disease subtypes | No. of ind. sig. SNPs | No. risk loci |
|---|---|---|
| CAD with T2DM | 9695 | 173 |
| CAD without T2DM | 2582 | 51 |
| T2DM without CAD | 7727 | 129 |
| CAD with Obesity | 599 | 22 |
| CAD without Obesity | 446 | 31 |
| Obesity without CAD | 314 | 17 |
| CAD with Stroke | 1189 | 34 |
| CAD without Stroke | 1871 | 33 |
| Stroke without CAD | 271 | 5 |
| T2DM with Obesity | 1744 | 69 |
| T2DM without Obesity | 2911 | 92 |
| Obesity without T2DM | 316 | 16 |
| T2DM with Stroke | 2585 | 72 |
| T2DM without Stroke | 12260 | 175 |
| Stroke without T2DM | 247 | 14 |
| Obesity with Stroke | 359 | 14 |
| Obesity without Stroke | 412 | 18 |
| Stroke without Obesity | 111 | 6 |

No. of ind. sig. SNPs indicates number of independent significant SNPs. Following the definition by the GWAS analytic platform FUMA, independent significant SNPs are defined as those that with p<5e-8 and are independent from each other at the default r2 threshold (r2=0.6). As for the definition of genomic loci, independent significant SNPs which are correlated with each other at $r2 \geq 0.1$ are assigned to the same risk locus.

*Heritability explained and genetic correlation among subtypes*

In order to uncover how much variance could be explained by all common variants in the GWAS panel, we calculated the SNP-based heritability of our studied disease combinations by LDSR[14] (Table 5). Interestingly, many comorbid cardiometabolic traits are more heritable than only single traits (without a comorbid disorder).

We also assessed the genetic correlation between different disease 'subtypes' as defined by the presence or absence of comorbid conditions. The results are summarized in Table 6. Interestingly, many pairs had weak or moderate genetic correlations, implying that they are possibly *distinct biological subtype*s of the disease. For example, comorbid CAD/T2DM only has a weak genetic correlation with non-diabetic CAD (rg = 0.111), suggesting that they may be genetically and biologically distinct 'subtypes'. Similarly, only a moderate correlation was observed between obese CAD and non-obese CAD (rg = 0.232). Furthermore, we compared the extent of overlap of significant genetic variants among different pairs of disease subtypes. As expected, the weaker the genetic correlation, the lesser the overlap of significant SNPs between disease subtypes.

*Genetic correlation and MR analysis*

To further explore the genetic overlap between the studied disease subtypes and other cardiometabolic conditions (mainly stroke/CAD), we analyzed their genetic correlations using LDSR (Table 7 and S9). This is also clinically relevant as we are often interested in whether a certain combination of traits is a significant risk factor for a certain disease. For example, do obese DM and non-obese DM confer the same risk to CAD? The findings will have implications for management and prevention of CAD. Here we performed LDSR and MR on several selected traits with higher clinical relevance (Table 7). For example, we observed that similar genetic correlation between obese and non-obese DM with CAD, suggesting the extent of genetic overlap with CAD are similar. On the other hand, the genetic correlation between obesity (without T2DM) per se and CAD is relatively weak (rg = 0.0797). As another example, while T2DM with obesity is moderately genetically correlated with stroke (rg = 0.2779), obesity without T2DM has *no* significant genetic correlation with stroke. We also assessed genetic correlations for a wider range of complex traits (856 traits in total) using LDSR; full results are shown in Table S9.

We then performed further MR analysis for selected pairs of traits to assess *causal* relationships between several disease subtypes and cardiovascular outcomes (Table 8 and S10). For simplicity, we primarily report the results at $r^2 = 0.05$, but most results are consistent across different $r^2$ thresholds. When focusing on CAD as the outcome, we found that obese DM is causally related to increased risk of CAD (MR-IVW; OR= 1.24, 95% CI: 1.16 to 1.33, p = 2.62E-11, Egger intercept p=0.355). Since we are studying binary exposures, the above OR roughly reflects the effect size with 2.72-fold increase in the exposure prevalence. Alternatively, the estimate may be multiple by 0.693 to reflect the OR resulted from doubling the prevalence of exposure[61], which is presented in our tables. Similar results were observed at other $r^2$ thresholds and with MR-Egger. Interestingly, we observed that *obesity without DM* does not have a significant causal link with CAD risks. On the other hand, non-obese DM showed no evidence of causal association under MR-Egger, but results were significant with MR-IVW. The Egger intercepts were significant (p<0.05 at all $r^2$ thresholds), suggesting that there is imbalanced horizontal pleiotropy, and that results from MR-Egger are more likely valid. The above finding suggests that some genetic variants may affect both non-obese DM and CAD risks via *different* pathways, leading to association between the two traits but the link may *not* be causal.

When considering stroke as the outcome, most disease subtypes studied were significantly and causally related to increased stroke risks. An exception is obesity alone without CAD or DM, which did not show a causal relationship with stroke. It is worthwhile to note that the effect size differs across different risk factors. For example, CAD comorbid with DM confers a higher risk (OR~ 1.12 per doubling of exposure prevalence; $r^2$ = 0.05) for stroke, compared to DM without CAD (OR~ 1.03 per doubling of exposure prevalence) or CAD without DM (OR ~ 1.06 per doubling of exposure prevalence).

Table 5 Heritability of 18 disease combinations from LD score regression on the observed scale

| Comorbidities | Heritability | Single Trait | Heritability | Single Trait | Heritability |
|---|---|---|---|---|---|
| CAD with T2DM | 0.0747 (0.0035) | CAD without T2DM | 0.0232 (0.0017) | T2DM without CAD | 0.0525 (0.0033) |
| CAD with Obesity | 0.1562 (0.0071) | CAD without Obesity | 0.1014 (0.0069) | Obesity without CAD | 0.1168 (0.0059) |
| CAD with Stroke | 0.0424 (0.0028) | CAD without Stroke | 0.0365 (0.0027) | Stroke without CAD | 0.0126 (0.0018) |
| T2DM with Obesity | 0.0681 (0.0027) | T2DM without Obesity | 0.0421 (0.0032) | Obesity without T2DM | 0.0241 (0.0014) |
| T2DM with Stroke | 0.0296 (0.0016) | T2DM without Stroke | 0.0568 (0.003) | Stroke without T2DM | 0.0077 (0.0009) |
| Obesity with Stroke | 0.0511 (0.0027) | Obesity without Stroke | 0.0574 (0.0028) | Stroke without Obesity | 0.0221 (0.0022) |

Table 6 Genetic correlation of different disease subtypes

| Subtype 1 | Subtype 2 | rg | P | No. overlapped sig. SNPs |
|---|---|---|---|---|
| T2DM with Obesity | T2DM without Obesity | 0.667 | **8.304E-170** | 335 |
| T2DM with Obesity | Obesity without T2DM | 0.5568 | **6.059E-108** | 290 |
| CAD with T2DM | CAD without T2DM | 0.111 | **0.0044** | 638 |
| CAD with T2DM | T2DM without CAD | 0.6888 | **6.12E-198** | 3333 |
| CAD with Obesity | CAD without Obesity | 0.2324 | **7.11E-12** | 115 |
| CAD with Obesity | Obesity without CAD | 0.5686 | **2.13E-158** | 274 |
| CAD with Stroke | CAD without Stroke | 0.9802 | **2.38E-153** | 325 |
| CAD with Stroke | Stroke without CAD | 0.3756 | **2.03E-12** | 117 |
| T2DM with Stroke | T2DM without Stroke | 0.9348 | **0.00E+00** | 1845 |
| T2DM with Stroke | Stroke without T2DM | 0.1343 | **1.57E-02** | 172 |
| Obesity with Stroke | Obesity without Stroke | 0.6711 | **4.36E-306** | 282 |
| Obesity with Stroke | Stroke without Obesity | 0.4096 | **7.6493E-21** | 9 |

Note: here rg indicates the genetic correlation between two traits. No. overlapped sig. SNPs indicates the number of SNPs with P < 5E-8 in both disease subtypes. Results with FDR<0.05 are in bold.

Table 7 Genetic correlations between studies disease subtypes and other cardiometabolic conditions

| Disease subtypes | Conditions | rg | P |
|---|---|---|---|
| T2DM with Obesity | CAD | 0.357 | **2.5714E-46** |
| T2DM without Obesity | CAD | 0.347 | **3.2467e-30** |
| Obesity without T2DM | CAD | 0.0797 | **0.0241** |
| CAD with T2DM | Stroke | 0.4737 | **9.1061E-39** |
| CAD without T2DM | Stroke | 0.2941 | **6.4598E-09** |
| T2DM without CAD | Stroke | 0.1368 | **0.0005** |
| T2DM with Obesity | Stroke | 0.2779 | **1.0613E-15** |
| T2DM without Obesity | Stroke | 0.3103 | **1.6109E-11** |
| Obesity without T2DM | Stroke | 0.0339 | 0.502 |
| CAD with Obesity | Stroke | 0.3595 | **1.509E-18** |
| CAD without Obesity | Stroke | 0.3585 | **1.45E-11** |
| Obesity without CAD | Stroke | -0.0126 | 0.7892 |

Note: here rg indicates the genetic correlation between two traits, gcov indicates the intercept for genetic covariance calculation, gcov_se indicates the standard error of gcov.

Table 8  MR analysis results of selected pairs of traits.

| Exposure | Outcome | Estimate | CILower | CIUpper | Pvalue | Method |
|---|---|---|---|---|---|---|
| T2DM without Obesity | CAD | 1.43E 03 | -6.61E-02 | 6.32E-02 | 9.65E-01 | Egger |
| CAD with T2DM | Stroke | 1.64E 01 | 1.29E-01 | 1.98E-01 | **3.36E-20** | IVW |
| T2DM with Obesity | CAD | 2.21E 01 | 1.56E-01 | 2.87E-01 | **2.62E-11** | IVW |
| T2DM without Obesity | Stroke | 7.77E 02 | 4.83E-02 | 1.07E-01 | **2.32E-07** | IVW |
| T2DM without CAD | Stroke | 4.63E 02 | 2.29E-02 | 6.97E-02 | **1.06E-04** | IVW |
| CAD without Obesity | Stroke | 1.20E 01 | 5.03E-02 | 1.90E-01 | **7.59E-04** | IVW |
| CAD with Obesity | Stroke | 1.02E 01 | 3.95E-02 | 1.65E-01 | **1.43E-03** | IVW |
| T2DM with Obesity | Stroke | 9.99E 02 | 3.63E-02 | 1.63E-01 | **2.07E-03** | IVW |
| CAD without T2DM | Stroke | 8.16E 02 | 2.15E-02 | 1.42E-01 | **7.82E-03** | IVW |
| Obesity without T2DM | CAD | 7.46E 02 | -2.96E-02 | 1.79E-01 | 1.60E-01 | IVW |
| Obesity without CAD | Stroke | 3.39E 02 | -1.85E-02 | 8.64E-02 | 2.05E-01 | IVW |
| Obesity without T2DM | Stroke | 3.99E 03 | -6.27E-02 | 7.07E-02 | 9.07E-01 | IVW |

Note: here CILower indicates the lower bound for the confidence interval, CIUpper indicates the upper bound for the confidence interval.

If the Egger intercept p-value was <0.05, the Egger regression approach was used; otherwise we employed the MR-IVW approach which has better statistical power.

**Extension to more than two traits**

As discussed above, our analytic framework may also be applied to the combination of three or more traits. To illustrate the methodology, we applied it to three cardiometabolic disorders, namely CAD, T2DM and obesity. Specifically, we explored the genetic architecture of obese T2DM comorbid with CAD, and non-obese T2DM with CAD. In brief, we applied the analytic method sequentially by first deriving the GWAS results of DM with and without obesity, then adding CAD as input in the next step.

Accordingly, we identified 76 and 91 genomic risk loci respectively for 'obese T2DM with CAD' and 'non-obese T2DM with CAD' that exceed genome-wide significance (Table S6). Details about gene and pathway analysis results were summarized in Table S7 and S8. Some genes/pathways are shared between the subtypes while some are top-ranked for specific subtypes. For a brief highlight of the results, for example, *TCF7L2* and *CDKN2B* were among the top susceptible genes for both disease subtypes, while *BDNF* and *HMGA2* were only found as risk genes in obese T2DM+CAD and non-obese T2DM+CAD respectively. As for pathways, plasma lipoprotein assembly, remodeling and clearance was one of the top enriched pathways unique to obese T2DM with CAD while anti-diabetic drug potassium channel inhibitors pathway was only found to be significantly enriched in the other disease subtype, i.e., non-obese T2DM with CAD.

**Application to clinically defined categories**

Furthermore, we applied our proposed framework to low-density lipoprotein cholesterol (LDL) based on publicly available summary statistics with a sample size of 188,577[62]. Typically, a LDL cholesterol level reading of 190 mg/dL or higher is considered as very high in clinical practice . Following this standard, we transformed the summary statistics of quantitative trait into that of binary trait. Then, we uncovered the genetic architectures of disease combinations involved CAD and high LDL utilizing our proposed framework. Totally, we identified 80, 40 and 65 genomic risk loci that exceed genome-wide significance respectively for CAD with high LDL, CAD without high LDL and high LDL without CAD. For details about these genomic risk loci, please refer to Table S6. Genes and pathways analysis results are shown in Table S7 and S8. The identified susceptibility genes and enriched pathways were linked to the pathophysiology of involved disorders[63-65]. As for tissue specificity analysis, we found that liver was the most significantly enriched tissue for CAD with high LDL.

**Real data replication**

We calculated the genetics correlations between GWAS summary statistics derived from real GWAS study and those estimated from our method for 3 different disease subtypes, i.e., CAD with T2DM, CAD without T2DM, and T2DM without CAD. Their genetic correlations are respectively 1.0181 (SE=0.0573) (CAD with T2DM), 0.8007 (SE=0.0306) (CAD without T2DM) and 1.021 (SE=0.0199) (T2DM without CAD). Obviously, they all achieved high genetic correlations, which further verified the validity of our proposed method.

**Discussion**

Here we have presented a statistical framework to uncover susceptibility variants for combination of diseases/traits, based on summary statistics alone. The method is useful for revealing the genetic basis of comorbid disorders, or disorders without relevant comorbidities. More broadly speaking, the cases can be considered as those affected by as specific 'subtype' of the disease (as characterized by the presence or absence of comorbid traits). We also extended the methodology to deal with continuous traits with clinically meaningful categories (e.g. lipid levels), and to more than 2 traits.

There are several strengths and potential applications of the proposed framework. Firstly, as the method only requires GWAS summary statistics, our framework can be potentially applied to a large variety of complex diseases. This approach is likely more cost-effective than recruiting subjects with comorbid disorders. As GWAS summary statistics with large sample sizes have dramatically increased these years, we believe the proposed framework may represent a new paradigm of analysis and will open up countless opportunities to study the genetic basis and architecture of disease combinations. Such efforts will help shed light on heterogeneity and pathophysiology of different complex disorders, and may contribute to the identification of new drug targets and more personalized therapies. Clinically, different 'subtypes' of a disease may be related to different complications. For example, we found that obese DM is causally related to increased risks of CAD, but obesity without DM may not be causal to development of CAD. Similarly, LDSR also suggested only a weak genetic correlation between obesity without DM and CAD. These analyses will help refine causal risk factors for diseases and the formulation of prevention strategies. Note that other secondary analysis of GWAS summary data, such as transcriptome-wide association studies (TWAS), Summary-data-based Mendelian Randomization (SMR; based on eQTLs, methylation QTLs etc.), other SNP-based (partitioned) heritability estimation, pathway analysis approaches etc. may also be readily applied although we only illustrated the application of several additional analyses.

There are a few limitations to the current study. Similar to other methodologies that employ summary statistics from more than one sample, such as two-sample MR, there is an implicit assumption that both sets of summary statistics (assuming the study of 2 traits) are based on the same population. Large heterogeneity between the samples (e.g. different ethnic groups, large differences in inclusion/exclusion criteria etc.) may lead to less accurate results. Also, most available summary statistics to date, including those we included in this study, are based on European samples. The results may not be transferable to other populations and it remains an open question how to accommodate summary data from different populations. While we believe the proposed framework is flexible and cost-effective, it could not completely replace the need to recruit subjects with comorbidities (or diseases without comorbidity). As discussed above, heterogeneity between study samples is inevitable, and recruitment of a more homogeneous sample with detailed phenotyping is still very valuable in uncovering the genetic basis of combination of diseases/traits. While we highlights several genes and pathways underlying cardiometabolic traits, further experimental studies are required to validate the findings.

Taken together, we believe the proposed approach is a useful extension to conventional single-trait analysis. Identification of genetic variants for comorbid disorders or disease 'subtypes' may ultimately lead to more targeted prevention and treatment, and identification of novel drug targets.

**Supplementary Materials are available at**

https://drive.google.com/open?id=1Q_FGQIslb5MY6pOx6_5lXy3Gi8CgQL2s

**Acknowledgements**

This work was supported partially by the Lo Kwee Seong Biomedical Research Fund, a Direct Grant from The Chinese University of Hong Kong and an NSFC Young Scientist grant (31900495). We are grateful to Prof. Stephen Tsui for computing support.

**Author contributions** Conceived and designed the study: HCS. Study supervision: HCS. Data analysis: LYY (lead), with input from CKLC, YPL and STR. Methodology: HCS (lead), LYY, with input from PCS. Data interpretation: HCS, LYY. Drafted the manuscript: LYY, HCS.

**Conflicts of interest**

The authors declare no conflict of interest.

**Figure legends**

Figure 1 Manhattan plot of GWAS results of comorbid disorders and single disorder without comorbidity